\documentclass[11pt]{article}
\usepackage[english]{babel}
\usepackage{latexsym}
\usepackage{fancybox}
\usepackage{pstricks}
\usepackage{pst-node}
\usepackage{pst-plot}

\long\def\COMMENT#1\ENDCOMMENT{\message{(Commented text...)}\par}

\def\ignore#1{}

\def\beq{\begin{equation}}
\def\eeq#1{\label{#1}\end{equation}}

\def\join#1{\mathrel{\mathop{
        \hbox{$\triangleright\!\triangleleft$}}
        \limits_{#1}}}

\def\beqa{\begin{eqnarray}}
\def\eeqa{\end{eqnarray}}

\def\Box{\hspace*{\fill}\framebox{\makebox[\depth]{\ }}}

\def\calS0{{\cal S}_0}

\newtheorem{definition}{Definition}[section]

\newtheorem{theorem}{Theorem}[section]

\newtheorem{proposition}{Proposition}[section]

\newtheorem{example}{Example}[section]

\def\wnnw#1{\omega_{#1}}
\def\wwnnw#1{\psi_{#1}}
\def\rnk#1{\eta_{#1}}

\title{{\bf Preference Queries}\thanks{This is an expanded version of the paper \cite{ChEDBT02}. CoRR paper cs.DB/0207093.}
}
\author{Jan Chomicki\\
Dept. of Computer Science and Engineering\\
University at Buffalo\\
Buffalo, NY 14260-2000\\
{\tt chomicki@cse.buffalo.edu}}

\begin{document}

\maketitle
\begin{abstract}
The handling of user preferences is becoming an increasingly
important issue in present-day information systems.
Among others, preferences are used for {\em information filtering and extraction}
to reduce the volume of data presented to the user.
They are also used to keep track of {\em user profiles} and formulate {\em policies}
to improve and automate decision making. 

We propose here a simple, logical framework for formulating preferences as {\em preference formulas}.
The framework does not impose any restrictions on the preference relations
and allows arbitrary operation and predicate signatures in preference formulas.
It also makes the composition of preference relations straightforward.
We propose  a simple, natural embedding of preference formulas into relational algebra
(and SQL) through a single {\em winnow} operator parameterized by a preference formula.
The embedding makes possible the formulation of complex preference queries,
e.g., involving aggregation, by piggybacking on existing SQL constructs.
It also leads in a natural way to the definition of further, preference-related
concepts like ranking.
Finally, we present general algebraic laws governing the winnow operator and its interaction
with other relational algebra operators.
The preconditions on the applicability of the laws are captured by logical formulas.
The laws provide a formal foundation for the algebraic optimization of preference queries.
We demonstrate the usefulness of our approach through numerous examples.
\end{abstract}

\section{Introduction}\label{s:intro}
The handling of user preferences is becoming an increasingly
important issue in present-day information systems.
Among others, preferences are used for {\em information filtering and extraction}
to reduce the volume of data presented to the user.
They are also used to keep track of {\em user profiles} and formulate {\em policies}
to improve and automate decision making. 

The research literature on preferences is extensive.
It encompasses preference logics \cite{VWr63,Man91,Han01},
preference reasoning \cite{WeDo91,TaPe94,BoBrHoPo99},
prioritized nonmonotonic  reasoning and logic programming
\cite{BrEi99,DeScTo00,SaIn00} and decision theory \cite{Fish99,Fish70}
(the list is by no means exhaustive).
However, only a few papers \cite{LaLa87,BoKoSt01,GoJaMa01,AgWi00,HrKoPa01,ChEDBT02,Kie02,KiKo02},
most of them very recent,
address the issue of user preferences in the context 
of database queries. Two different approaches are pursued:
qualitative and quantitative.
In the {\em qualitative} approach \cite{LaLa87,BoKoSt01,GoJaMa01,ChEDBT02,Kie02,KiKo02}, the preferences
between tuples in the answer to a query are specified directly,
typically using binary {\em preference relations}.
\begin{example}\label{ex:sal}
We introduce here one of the examples used throughout the paper.
Consider the relation $Book(ISBN,Vendor,Price)$
and the following preference relation $\succ_1$ between {\em Book} tuples:
\begin{quote}
{\em prefer one Book tuple to another if and only if their ISBNs are the same and the Price of the
first is lower.}
\ignore{
{\em if two tuples have the same ISBN and different Price, prefer the one with the lower Price}.}
\end{quote}
Consider the following instance $r_1$ of $Book$

\begin{center}
\begin{tabular}{|l|l|l|}
\hline
{\em ISBN} &{\em Vendor} &{\em Price}\\\hline
0679726691 & BooksForLess & \$14.75\\
0679726691 & LowestPrices & \$13.50\\
0679726691 & QualityBooks & \$18.80\\
0062059041 & BooksForLess & \$7.30\\
0374164770 & LowestPrices & \$21.88\\\hline
\end{tabular}
\end{center}

Then clearly the second tuple is preferred to the first one which in turn
is preferred to the third one. There is no preference defined between
any of those three tuples and the remaining tuples.
\end{example}
In the  {\em quantitative} approach \cite{AgWi00,HrKoPa01}, preferences are 
specified indirectly using {\em scoring functions} that associate 
a numeric score with every tuple of the query answer.
Then a tuple $t_1$ is preferred to a tuple $t_2$ iff the score
of $t_1$ is higher than the score of $t_2$.
The qualitative approach is strictly more general than the quantitative one,
since one can define preference relations in terms of scoring functions
(if the latter are explicitly given), while not every intuitively plausible
preference relation can be captured by scoring functions.
\begin{example}\label{ex:sal:01}
There is no scoring
function that captures the preference relation described in Example \ref{ex:sal}.
Since there is no preference defined between any of the first three tuples
and the fourth one, the score of the fourth tuple should be equal to all of 
the scores of the first three tuples. But this implies that the scores
of the first three tuples are the same, which is not possible since 
the second tuple is preferred to the first one which in turn is preferred
to the third one.
\end{example}
This lack of expressiveness of the quantitative approach is well known
in utility theory \cite{Fish99,Fish70}.

In the present paper, we contribute to the qualitative approach by
defining a logical framework for formulating preferences and its embedding
into relational query languages.

\ignore{
Specifying preferences in user queries,
instead of associating them with database objects has the following advantages:
\begin{itemize}
\item {\em flexibility:} different users can apply different preferences
to the same database. For example, the database may contain 
restaurant menus, and the users can query it stating their own personal
meal preferences.
\item {\em amenability to query optimization:} since the preferences
are specified in queries, they will be optimized together with other
query constructs.
\end{itemize}
}

We believe that combining preferences
with queries is very natural and useful. The applications in which
user preferences are prominent will benefit from applying the modern
database technology. For example, in decision-making applications
databases may be used to store the space of possible configurations.
Also, the use of a full-fledged query language makes it possible to 
formulate complex decision problems, a feature missing from most previous,
non-database, approaches to preferences. For example, the formulation of the problem
may now involve quantifiers, grouping, or aggregation.
At the same time by explicitly
addressing the technical issues involved in querying with
preferences present-day DBMS may expand their scope. 

The framework presented in this paper consists of two parts:
a formal first-order logic notation for specifying preferences
and an embedding of preferences into relational query languages.
In this way both abstract properties of preferences (like
asymmetry or transitivity) and evaluation of preference queries can be studied
to a large degree separately.

Preferences are defined using binary {\em preference relations} between
tuples. Preference relations are specified using first-order formulas.
We focus mostly on {\em intrinsic} preference formulas. Such formulas can refer only 
to built-in predicates. In that way we capture preferences that are
based only on the values occuring in tuples, not on other properties
like membership of tuples in database relations. We show how the latter kind
of preferences, called {\em extrinsic}, 
can also be simulated in our framework in some cases.

We propose a new relational algebra operator called {\em winnow} that
selects from its argument relation the {\em most preferred tuples} according
to the given preference relation. Although the winnow operator can
be expressed using other operators of relational algebra, by 
considering it on its own we can on one hand focus on the abstract properties of
preference relations (e.g., transitivity) and on the other, study special
evaluation and optimization techniques for the winnow operator itself.
For SQL. we are faced with a similar choice: 
either the language is appropriately extended with an SQL equivalent of winnow,
or the occurrences of winnow are translated into SQL.
The first alternative looks more promising; however, in this paper we don't
commit ourselves to any specific syntactic expression of winnow in SQL.

We want to capture many different varieties of preference and related notions:
{\em unconditional\/} vs. {\em conditional\/} preferences,
{\em nested\/} and {\em hierarchical\/} preferences,
{\em groupwise\/} preferences,
{\em indifference\/},
{\em iterated} preferences and {\em ranking}, and
{\em integrity constraints\/} and {\em vetoes\/}.

The main contributions of this paper are as follows:
\begin{enumerate}
\item a simple, logical framework for formulating preferences as preference formulas.
The framework does not impose any restrictions on the preference relations
and allows arbitrary operation and predicate signatures in preference formulas.
It also makes the composition of preference relations straightforward.
\item a simple, natural embedding of preference formulas into relational algebra
(and SQL) through a single winnow operator parameterized by a preference formula.
The embedding makes possible the formulation of complex preference queries,
e.g., involving aggregation, by piggybacking on existing SQL constructs.
It also leads in a natural way to the definition of further, preference-related
concepts like ranking.
\item general algebraic laws governing the winnow operator and its interaction
with other relational algebra operators.
The preconditions on the applicability of the laws are captured by logical formulas.
The laws provide a formal foundation for the algebraic optimization of preference queries.
\end{enumerate}
 
In Section \ref{sec:basic}, we define the basic concepts of preference
relation, preference formula, and the winnow operator.
We also introduce several examples that will be used throughout the paper.
In Section \ref{sec:prop}, we study the basic properties of preference relations.
In Section \ref{sec:winnow}, which contains the main technical contributions of the
paper, we present the main properties of the winnow operator, characterize its
expressive power, and outline -- for completeness -- a number of evaluation algorithms that were
proposed elsewhere.
In Section \ref{sec:spec}, we explore the composition of preferences.
In Section \ref{sec:ext}, we show how the winnow operator together
with other constructs of relational algebra and SQL makes it possible 
to express  a wide variety of preference queries.
In Section \ref{sec:iter}, we show how iterating the winnow
operator provides a ranking of tuples and introduce a weak version
of the winnow operator that is helpful for  preference
relations that are not strict partial orders.
We discuss related work in Section \ref{sec:related} and conclude
with a brief discussion of further work in Section \ref{sec:concl}.
All the non-trivial proofs are given.

\section{Basic notions}\label{sec:basic}

We are working in the context of the relational model of data.
We assume two infinite domains: $D$ (uninterpreted constants) and $N$ (numbers).
We do not distinguish between different numeric domains, since
it is not necessary for the present paper.
When necessary, we assume that database instances are finite. 
(Some results hold without the finiteness assumption.)
Additionally,
we have the standard built-in predicates.
In the paper, we will move freely between relational algebra and SQL.

\subsection{Basic definitions}
Preference formulas are used to define binary preference relations.

\begin{definition}\label{def:prefrel}
Given a relation schema $R(A_1 \cdots A_k)$ 
such that $U_i$, $1\leq i\leq k$, is the domain (either $D$ or $N$)
of the attribute $A_i$, a relation $\succ$ is a {\em preference relation over $R$}
if it is a subset of $(U_1\times\cdots\times U_k)\times (U_1\times\cdots\times U_k)$.
\end{definition}

Intuitively, $\succ$ will be a binary relation between pairs of tuples from the
same (database) relation. 
We say that a tuple $t_1$ {\em dominates} a tuple $t_2$
in $\succ$ if $t_1\succ t_2$. 

Typical properties of the relation $\succ$ include:
\begin{itemize}
\item {\em irreflexivity}: $\forall x. x\not\succ x,$
\item {\em asymmetry}: $\forall x,y. x\succ y\Rightarrow y\not\succ x,$
\item {\em transitivity}: $\forall x,y,z. (x\succ y \wedge y\succ z)\Rightarrow x\succ z,$
\item {\em negative transitivity}: $\forall x,y,z. (x\not\succ y \wedge y\not\succ z)\Rightarrow x\not\succ z,$
\item {\em connectivity}: $\forall x,y. x\succ y\vee y\succ x \vee x=y.$
\end{itemize}
 
The relation $\succ$ is: a {\em strict partial order} if it is
irreflexive, asymmetric and transitive; a {\em total order} if it is
a connected strict partial order; a {\em weak order} if it is is a 
negatively transitive strict partial order.
At this point, we do not assume any properties
of $\succ$, although in most applications it will satisfy at least the properties of a strict partial
order.

\begin{definition}\label{def:prefformula}
A {\em preference formula (pf)} $C(t_1,t_2)$ is a first-order formula defining 
a preference relation $\succ_C$ in the standard sense, namely
\[t_1\succ_Ct_2\;{\rm iff}\; C(t_1,t_2).\]
An {\em intrinsic preference formula (ipf)} is a preference formula that
uses only built-in predicates.
\end{definition}

We will limit our attention to preference relations defined using preference
formulas.
By using the notation $\succ_C$ for a preference relation, we assume that there is 
an underlying preference formula $C$.

Ipfs can refer to equality ($=$) and inequality ($\not=$) when
comparing values that are uninterpreted constants, and to the standard
set of built-in arithmetic comparison operators when referring to
numeric values (there are no function symbols). 
We will call an ipf that references only arithmetic comparisons
$(=,\not=,<,>,\leq,\geq)$ {\em pure comparison}.
Without loss of generality, we
will assume that ipfs are in DNF (Disjunctive Normal Form) and
quantifier-free (the theories involving the above predicates admit
quantifier elimination).
A formula in DNF is called $k$-DNF if it has at most $k$ disjuncts.

In this paper, we mostly restrict ourselves to ipfs and preference relations
defined by such formulas.
The main reason is that ipfs define {\em fixed}, although possibly infinite,
relations. As a result, they
are computationally easier and more amenable
to syntactic manipulation that general pfs.
For instance, transitively closing an ipf results in a finite formula
(Theorem \ref{th:trans}), which is typically not the case for pfs.
However, we formulate in full generality the results that hold for
arbitrary pfs.

We define now an algebraic operator that picks from a given relation the
set of the {\em most preferred tuples}, according to a given preference formula.
\begin{definition}\label{def:winnow}
If $R$ is a relation schema and $C$ a preference
formula defining a preference relation $\succ_C$ over $R$,
then the {\em winnow operator} is written as $\wnnw{C}(R)$, 
and for every instance $r$ of $R$:
\[\wnnw{C}(r)=\{t\in r\mid\neg \exists t'\in r.\ t'\succ_C t\}.\]
\end{definition}

A preference query is a relational algebra query containing at least
one occurrence of the winnow operator.

\subsection{Examples}

The first example illustrates how preference queries are
applied to {\em information extraction}: here obtaining
the best price of a given book.

\begin{example}\label{ex:sal:02}
Consider the relation $Book(ISBN,Vendor,Price)$
from Example \ref{ex:sal}. 
The preference relation $\succ_{C_1}$ from this example
can be defined using the formula $C_1$:
\[(i,v,p)\succ_{C_1}(i',v',p') \equiv i=i' \wedge p<p'.\]
The answer to the preference query 
$\wnnw{C_1}(Book)$
provides for every book the information about the vendors offering the 
lowest price for that book.
For the given  instance $r_1$ of $Book$,
applying the winnow operator $\wnnw{C_1}$ 
returns the tuples
\begin{center}
\begin{tabular}{|l|l|l|}
\hline
{\em ISBN} &{\em Vendor} &{\em Price}\\\hline
0679726691 & LowestPrices & \$13.50\\
0062059041 & BooksForLess & \$7.30\\
0374164770 & LowestPrices & \$21.88\\\hline
\end{tabular}
\end{center}

Note that in the above example, the preferences are applied
{\em groupwise}: separately for each book.
Note also that due to the properties of $<$, the preference relation
$\succ_{C_1}$ is irreflexive, asymmetric and transitive.
\end{example}

The second example illustrates how preference queries are used
in {\em automated decision making} to obtain the most 
desirable solution to a (very simple) configuration problem.

\begin{example}\label{ex:wine}
Consider two relations $Wine(Name,Type)$ and $Dish(Name,Type)$
and a view $Meal$ that contains possible meal configurations
\begin{verbatim}
CREATE VIEW Meal(Dish,DishType,Wine,WineType) AS
  SELECT * FROM Wine, Dish;
\end{verbatim}
Now the preference for white wine in the presence
of fish and for red wine in the presence of meat
can be expressed as the following preference formula $C_2$ over $Meal$:
\[\begin{array}{lcl}
(d,dt,w,wt)\succ_{C_2} (d',dt',w',wt') &\equiv& (d=d' \wedge dt={\rm 'fish'}\wedge wt={\rm 'white'} \\
&&\wedge dt'={\rm 'fish'} \wedge wt'={\rm 'red'})\\
&&\vee (d=d'\wedge dt={\rm 'meat'}\wedge wt={\rm 'red'} \\
&&\wedge dt'={\rm 'meat'}\wedge wt'={\rm 'white'})
\end{array}\]
Notice that this will force any white wine to be preferred over any
red wine for fish, and just the opposite for meat.
For other kinds of dishes, no preference is indicated.
This is an example of a {\em relative} preference.
Consider now the preference query $\wnnw{C_2}(Meal)$.
It will pick the most preferred meals, according to the above-stated preferences.
Notice that in the absence of any white wine, red wine
can be selected for fish.

The above preferences are conditional, since they depend on the 
type of the dish being considered.
Note that the relation $\succ_{C_2}$ in this example is irreflexive and asymmetric.
Transitivity is obtained trivially because the chains of $\succ_{C_2}$ are
of length at most 2.
Note also that the preference relation is defined without referring
to any domain order.

Note also that the meals with a wine which is neither red nor white but, e.g., ros\'{e},
are not related through $\succ_{C_2}$ to the meals with either of those kinds of wine.
Therefore, the preference query $\wnnw{C_2}(Meal)$ will return also the meals
involving such wines, as they are not dominated by other meals. If this is undesirable,
one can express an absolute preference for white wine for fish (and red wine for meat)
using the formula $C_3$:
\[\begin{array}{lcl}
(d,dt,w,wt)\succ_{C_3} (d',dt',w',wt') &\equiv& (d=d' \wedge dt={\rm 'fish'}\wedge wt={\rm 'white'} \\
&&\wedge dt'={\rm 'fish'} \wedge wt'\not={\rm 'white'})\\
&&\vee (d=d'\wedge dt={\rm 'meat'}\wedge wt={\rm 'red'} \\
&&\wedge dt'={\rm 'meat'}\wedge wt'\not={\rm 'red'})
\end{array}\]
Similarly, an unconditional preference for red wine for any kind of meal can also
be defined as a first-order formula $C_4$:
\[\begin{array}{lcl}
(d,dt,w,wt)\succ_{C_4} (d',dt',w',wt') &\equiv& d=d'\wedge wt={\rm 'red'} \wedge wt'\not={\rm 'red'}.
\end{array}\]
\end{example}

\section{Properties of preference queries}\label{sec:prop}

\subsection{Preference relations}

Since pfs can be essentially arbitrary formulas, no properties of preference relations
can be assumed. So our framework is entirely neutral in this respect.

In the examples above, the preference relations were strict partial orders.  
This is likely to be the case for most applications of preference
queries. However, there are cases where such relations fail to satisfy
one of the properties of partial orders. 
We will see in Section \ref{sec:ext} when
irreflexivity fails.
For asymmetry:
We may have two tuples $t_1$ and $t_2$ such that $t_1\succ t_2$ and
$t_2\succ t_1$ simply because we may have one reason to prefer $t_1$
over $t_2$ and another reason to prefer $t_2$ over $t_1$.
Similarly, transitivity is not always guaranteed \cite{Hug80,Man91,Fish99,Han01}.
For example, $t_1$ may be
preferred over $t_2$ and $t_2$ over $t_3$, but the gap between
$t_1$ and $t_3$ with respect to some heretofore ignored property
may be so large as to prevent preferring $t_1$ over $t_3$.
Or, transitivity may have to be abandoned to prevent cycles in preferences.
\ignore{
As we show later (Theorem ???), transitivity is actually irrelevant from the 
semantic point of view, since transitively closing a preference relation $\succ_C$
leaves the result of the query $\wnnw{C}(r)$ invariant.
}
However, transitivity is essential for the correctness of the algorithms that
compute winnow (Section \ref{sec:winnow}).

It is not difficult to check the properties of a preference
relation defined using a pure comparison ipf.

\begin{theorem}\label{th:asym}
If a preference relation is defined using
a pure comparison ipf in DNF, it can be checked in PTIME for irreflexivity and asymmetry.
If the ipf is also in $k$-DNF for some fixed $k$, then the preference relation
can be checked in PTIME for transitivity, negative transitivity, and connectivity.
\end{theorem}
{\bf Proof:}
We discuss first asymmetry, the remaining  properties can be handled in a similar way.
If $t_1\succ t_2$ is defined as $D_1\vee\ldots\vee D_m$
and $t_2\succ t_1$ as $D_1'\vee\ldots\vee D_m'$, we
can write down the negation of asymmetry as 
$(D_1\vee\ldots\vee D_m)\wedge(D_1'\vee\ldots\vee D_m')$.
This formula is satisfiable iff at least one of $m^2$ formulas $\phi_{i,j}\equiv
D_i\wedge D_j'$, $i,j=1,\ldots,m$, is satisfiable.
Each formula $\phi_{i,j}$ is a conjunction of atomic formulas involving arithmetic
comparison predicates. Thus its satisfiability can be checked
in PTIME using the methods of \cite{GuSuWe96}.
Testing for transitivity, negative transitivity and connectivity requires writing down the negation of a DNF
formula and distributing the negation inside. 
The restriction to $k$-DNF guarantees that we have again a polynomial number of
PTIME satisfiability problems.
\hfill $\Box$

\begin{theorem}\label{th:nonempty}
If a preference relation $\succ_C$ over $R$ is a strict partial order,
then for every finite, nonempty instance $r$ of $R$, $\wnnw{C}(r)$ is nonempty.
\end{theorem}

If the properties of strict partial orders are not satisfied, then Theorem
\ref{th:nonempty} may fail to hold and the winnow operator may return an
empty set, even though the relation to which it is applied is nonempty.
For instance, if $r_0=\{t_0\}$ and $t_0\succ t_0$ 
(violation of irreflexivity), then the winnow
operator applied to $r_0$ returns an empty set.
Similarly, if two tuples are involved in a violation of asymmetry, they
may block each other from appearing in the result of the winnow operator.
Also, if the relation $r$ is infinite, it may happen that $\wnnw{C}(r)=\emptyset$,
for example if $r$ contains all natural numbers and the preference relation
is the standard ordering $>$.
 
The winnow operator is not {\em monotone} or {\em anti-monotone}.
\begin{example}
Consider the following preference formula $C_6$:
\[x\succ_{C_6} y \equiv x=a \wedge y=b.\]
Then
\[b=\wnnw{C_6}(\{b\})\not\subseteq\wnnw{C_6}(\{a,b\})=a.\]
Thus monotonicity and anti-monotonicity fail.
\end{example}

However, a form of monotonicity with respect to the preference formula parameter holds 
for winnow.
\begin{theorem}\label{th:monotone}
If $\succ_{C_1}$ and $\succ_{C_2}$ are preference relations over a relation schema $R$, and the formula 
\[\forall t_1,t_2[C_1(t_1,t_2)\Rightarrow C_2(t_1,t_2)]\]
is valid, then for all instances $r$ of $R$,
$\wnnw{C_2}(r)\subseteq\wnnw{C_1}(r)$.
If $\succ_{C_1}$ and $\succ_{C_2}$ are strict partial orders, then the converse also holds.
\end{theorem}
{\bf Proof:}
The first part is obvious. To see that the second part also holds,
assume that for all relations $r$,
$\wnnw{C_2}(r)\subseteq\wnnw{C_1}(r)$ but $C_1\not\Rightarrow C_2$.
Thus, $C_1 \wedge \neg C_2$ is satisfiable, and there are two
tuples $t_1$ and $t_2$ such $t_1\succ_{C_1} t_2$ but $t_1\not\succ_{C_2} t_2$.
Consider now the instance $r_{12}=\{t_1,t_2\}$.
Then $\wnnw{C_1}(r_{12})=\{t_1\}$ but $t_2\in\wnnw{C_2}(r_{12})$, a contradiction.
\hfill $\Box$

\ignore{
In addition to the properties of partial orders, 
an important property of preference relations is {\em acyclicity}.
It may be the case that $t_1\succ t_2$, $t_2\succ t_3$ but $t_3\succ t_1$.
In this case the tuples $t_1$, $t_2$ and $t_3$ will prevent one another
from being selected by the winnow operator. 
If there are no tuples that dominate them, the result of applying the winnow
operator may be empty.
We discuss the computational issues associated with acyclicity
in the next section.}

Several properties of winnow follow directly from the definition (the first is listed
in \cite{Kie02}, although in a less general context):
\begin{proposition}
For every preference relations $\succ_{C_1}$ and $\succ_{C_2}$ over a relation
schema $R$ and every instance $r$ of $R$:
\[\begin{array}{l}
\wnnw{C_1\vee C_2}(r)=\wnnw{C_1}(r)\cap\wnnw{C_2}(r)\\
\wnnw{False}(r)=r\\
\wnnw{True}(r)=\emptyset.\end{array}\]
\end{proposition}

\subsection{Indifference}

There is a natural notion of {\em indifference} 
associated with our approach: two tuples $t_1$ and $t_2$ are {\em indifferent}
($t_1\sim_C t_2$) if
neither is preferred to the other one, i.e.,
$t_1\not\succ_C t_2$ and $t_2\not\succ_C t_1$.
\begin{proposition}
For every preference relation $\succ_C$, every relation $r$ and every tuple $t_1,t_2\in\wnnw{C}(r)$,
we have $t_1=t_2$ or $t_1\sim_C t_2$.
\end{proposition}

It is a well-known result in decision theory \cite{Fish99,Fish70} that 
in order for a preference relation to be representable using scoring functions
the relation has to be a weak order. This implies, in particular, that 
the corresponding indifference relation (defined as above) has to be
{\em transitive}. This is not the case for the preference relation
$\succ_{C_1}$ defined in Example \ref{ex:sal}.

\section{The winnow operator}\label{sec:winnow}
In this section, we study various properties of the winnow operator:
expressive power, monotonicity, commutativity and distributivity.
Formulating such properties is essential for the evaluation and optimization
of preference queries. We also briefly discuss some evaluation methods
for winnow.

Although, as we show, the winnow operator can be expressed in relational algebra,
its explicit use makes possible a clean separation of preference
formulas from other aspects of the query.
This has several advantages.
First, the properties of preference relations can be studied
in an abstract way.
Second, specialized query evaluation methods for the winnow operator
can be developed.
Third, algebraic properties of that operator can be formulated, in order
to be used in query optimization.

\subsection{Expressive power}
The winnow operator can be expressed in relational algebra, and thus does not add any expressive
power to it. Perhaps more surprisingly, winnow can be used to simulate set difference.

\begin{theorem}\label{th:RA}
Relational algebra with winnow replacing set difference has the same expressive power
as standard relational algebra.
\end{theorem}
{\bf Proof:}
Clearly, the winnow operator is first-order definable. Thus any relational algebra query
with winnow can be translated to relational calculus, and then back to relational
algebra (without winnow).
Such a construction is, however, mainly of theoretical importance.

From a practical point of view, we show now the translation of the winnow operator
$\wnnw{C}(r)$ for $C=D_1\vee\ldots\vee D_k$ which is a pure comparison ipf formula in DNF.
Each $D_i$, $i=1,\ldots,k$, is a formula over free
variables $t_1$ and $t_2$.
It can be viewed as a conjunction
$D_i\equiv \phi_i\wedge\psi_i\wedge\gamma_i$ where $\phi_i$ refers only to the
variables of $t_1$, $\psi_i$ to the variables of $t_2$,
and $\gamma_i$ to the variables of both $t_1$ and $t_2$.
The formula $\phi_i$ has an obvious translation to a selection condition 
$\Phi_i$ over $R$, and the formula $\psi_i$ a similar translation
to a selection condition $\Psi_i$ over $\varrho(R)$, where $\varrho$ is a renaming of $R$.
The formula $\gamma_i$ can similarly be translated to a join condition
$\Gamma_i$ over $R$ and $\varrho(R)$.
Then 
\[\wnnw{C}(R)=\varrho^{-1}(\varrho(R)-\pi_{\varrho(R)}(\bigcup_{i=1}^k (\sigma_{\Phi_i}(R) 
\join{\Gamma_i}\sigma_{\Psi_i}(\varrho(R)))))\]
where $\varrho^{-1}$ is the inverse of the renaming $\varrho$.

We show now how to simulate the set difference operator $R-S$ using winnow.
Assume that $R$ (and $S$) have the set of attributes $X$ of arity $k$.
Then
\[R-S=\pi_{X}(\sigma_{B\not= 0}(\wnnw{C_5}(R\times \{1\}\cup S\times\{0\})))\]
where $B$ is the last attribute of $R\times \{1\}$ and
\[(x_1,\ldots,x_k,b)\succ_{C_5}(x_1',\ldots,x_k',b')\equiv x_1=x_1'\wedge\cdots \wedge x_k=x_k'\wedge b=0 \wedge b'=1.\]
This works as follows. Think of the attribute $B$ as a tag.
All the tuples in $R$ (resp. $S$) are tagged with $1$ (resp. $0$).
If a tuple is in $R\cap S$, then there are two copies of it
in $R\times \{1\}\cup S\times\{0\}$: one tagged with $1$, the other with $0$.
The latter one is preferred according to $\succ_{C_5}$. Finally, the selection $\sigma_{B\not=0}$ eliminates
all the tuples in $S$, keeping the tuples that are only in $R$.
\hfill $\Box$

\subsection{Evaluating winnow}
For completeness, we show here several algorithms that can be used to compute the result of the winnow
operator $\wnnw{C}(r)$.
The first is a simple nested-loops algorithm (Figure \ref{fig:NL}).
The second is BNL, an algorithm proposed in \cite{BoKoSt01} in the context
of {\em skyline queries}, a specific class of preference queries,
but the algorithm is considerably more general (Figure \ref{fig:BNL}).
The third \cite{ChGoGrLi02} is a variant of the second, in which a presorting step is used (Figure \ref{fig:SFS}).
All the algorithms used a fixed amount of main memory (a {\em window}).
However, for the algorithm NL, this is not made explicit, since it is irrelevant
for the properties of the algorithm that are of interest here.
Our emphasis  is not on the algorithms themselves -- they are much more completely
described and analyzed in the original papers -- but rather on determining their
scope. We will identify the classes of preference queries to which each of them
is applicable.

\begin{small}
\begin{figure}[hf]
\centering
\fbox{%
\begin{minipage}{.8\textwidth}
\begin{small}
\begin{enumerate}
\item open a scan $S_1$ on $r$;
\item for every tuple $t_1$ returned by $S_1$:
\begin{enumerate}
\item open a scan $S_2$ on $r$;
\item for every tuple $t_2$ returned by $S_1$:
\begin{tabbing}
if $t_2\succ_C t_1$, then close $S_2$ and goto \ref{label:close};
\end{tabbing}
\item output $t_1$; 
\item\label{label:close} close $S_2$;
\end{enumerate}
\item close $S_1$.
\end{enumerate}
\caption{NL: Nested Loops}
\label{fig:NL}
\end{small}\end{minipage}}
\end{figure}

\begin{figure}[hf]
\centering
\fbox{%
\begin{minipage}{.8\textwidth}
\begin{small}\begin{enumerate}
\item initialize the window $W$ and the temporary table $F$ to empty;
\item make $r$ the input;
\item repeat the following until the input is empty:
\begin{enumerate}
\item for every tuple $t$ :
\begin{itemize}
\item $t$ is dominated by a tuple in $W$ $\Rightarrow$ ignore $t$,
\item $t$ dominates some tuples in $W$ $\Rightarrow$ eliminate the dominated tuples
and insert $t$ into $W$,
\item $t$ is incomparable with all tuples in $W$ $\Rightarrow$ insert $t$ into $W$
(if there is room), otherwise add $t$ to $F$;
\end{itemize}
\item output the tuples from $W$ that were added there when $F$ was empty,
\item make $F$ the input, clear $F$.
\end{enumerate}
\end{enumerate}
\caption{BNL: Blocked Nested Loops}
\label{fig:BNL}
\end{small}\end{minipage}}
\end{figure}

\begin{figure}[hf]
\centering
\fbox{%
\begin{minipage}{.8\textwidth}
\begin{small}
\begin{enumerate}
\item topologically sort $r$ according to $\succ_C$;
\item make $r$ the input;
\item initialize the window $W$ and the temporary table $F$ to empty;
\item repeat the following until the input is empty:
\begin{enumerate}
\item for every tuple $t$ in the input:
\begin{itemize}
\item $t$ is dominated by a tuple in $W$ $\Rightarrow$ ignore $t$,
\item $t$ is incomparable with all tuples in $W$ $\Rightarrow$ insert $t$ into $W$
(if there is room), otherwise add $t$ to $F$;
\end{itemize}
\item output the tuples from $W$.
\item make $F$ the input, clear $F$.
\end{enumerate}
\end{enumerate}
\caption{SFS: Sort-Filter-Skyline}
\label{fig:SFS}
\end{small}\end{minipage}}
\end{figure}
\end{small}
The NL algorithm is correct for any preference relation $\succ_C$.
In principle, the preference relation might even be reflexive, since
the algorithm compares a tuple with itself.
The BNL and SFS algorithms require the preference relation to be a strict
partial order (for BNL this is noted in \cite{BoKoSt01}). The algorithms require irreflexivity, because they
do not compare a tuple with itself. Neither do they
handle correctly symmetry: the situation where there are two tuples $t_1$ and $t_2$
such that $t_1\succ_C t_2$ and $t_2\succ_C t_1$. In this case, BNL
will break the tie depending on the order in which the tuples appear, and SFS
will fail altogether, being unable to produce a topological sort.
To see the necessity of transitivity, consider the following example.
\begin{example}\label{ex:BNL:trans}
The preference relation $C_0$ is defined as follows:
\[x\succ_{C_0} y\equiv x=a\wedge y=b\vee x=b\wedge y=c.\]
Now let us suppose that the window has room for only one tuple,
and the tuples arrive in the following order: $a$, $b$, $c$.
Then $a$ will be in the window, and $b$ will be discarded, which prevents $b$ from 
blocking $c$. Therefore, BNL will output $a$ (correctly) and $c$ (incorrectly).
Such an example can be easily generalized to any fixed window size, simply be assuming
that $a$ and $b$ are separated in the input by sufficiently many values different from $a$, $b$ and $c$.
\end{example}

\subsection{Algebraic laws}

We present here a set of algebraic laws that govern the commutativity
and distributivity of winnow w.r.t. relational algebra operators. 
This set constitutes a formal foundation for rewriting preference queries
using the standard strategies like {\em pushing selections down}.
We prove the soundness of the introduced laws. In the cases of selection, projection, union and difference, we show
that the preconditions on the applicability of the laws are not only sufficient
but also necessary.
In the remaining cases, we show that the violations of the preconditions lead to the
violations of the laws.
In most interesting cases, the preconditions can also be efficiently checked.
\ignore{
We also argue for their completeness
in the sense that the conditions on their applicability cannot be weakened.
It is unclear how to obtain any stronger notion of completeness in this
context, since, to our knowledge, such a notion has not been established even
for the standard relational algebra without the winnow operator.
}

We adopt the set-based view of relational algebra operators and leave exploring
the multiset-based view for future research.

\subsubsection{Commutativity of winnow}

We establish here a sufficient condition for winnow to be commutative.
Commutativity is a fundamental property that makes it possible to move
the winnow operator around in preference queries.

\begin{theorem}\label{th:commute}
If $C_1$ and $C_2$ are preference formulas over a schema $R$ such that 
\begin{itemize}
\item the formula $\forall t_1,t_2[C_1(t_1,t_2)\Rightarrow C_2(t_1,t_2)]$
is valid, and
\item $\succ_{C_1}$ and $\succ_{C_2}$ are strict partial orders, 
\end{itemize}
then for all finite instances $r$ of $R$:
\[\wnnw{C_1}(\wnnw{C_2}(r))=\wnnw{C_2}(\wnnw{C_1}(r))=\wnnw{C_2}(r).\]
\end{theorem}
{\bf Proof:}
We prove here the first equality; the second can be proved in a similar way.

Assume $t\notin\wnnw{C_2}(\wnnw{C_1}(r))$ and $t\in \wnnw{C_1}(\wnnw{C_2}(r))$.
Then also $t\in\wnnw{C_2}(r)$. There are two possibilities:
(1) $\exists t'\in\wnnw{C_1}(r)$ such that $t'\succ_{C_2} t$. But then
$t'\in r$, which contradicts the fact
that $t\in\wnnw{C_2}(r)$.
(2) $t\notin\wnnw{C_1}(r)$. But then by Theorem \ref{th:monotone}
$t\notin\wnnw{C_2}(r)$, a contradiction.

Assume $t\notin\wnnw{C_1}(\wnnw{C_2}(r))$ and $t\in \wnnw{C_2}(\wnnw{C_1}(r))$.
Then also $t\in\wnnw{C_1}(r)$. There are two possibilities:
(1) $\exists t'\in\wnnw{C_2}(r)$ such that $t'\succ_{C_1} t$. But then also
$t'\in r$, which contradicts the fact
that $t\in\wnnw{C_1}(r)$.
(2) $t\notin\wnnw{C_2}(r)$. Still $t\in r$, since otherwise $t\notin\wnnw{C_1}(r)$.
Therefore, $\exists t'\in r$ such that $t'\succ_{C_2}t$. Now because $\succ_{C_2}$ 
is a strict partial order and $r$ is finite, we can choose $t'\in\wnnw{C_2}(r)$.
If  $t'\in\wnnw{C_1}(r)$, then in view of the fact that $t\in\wnnw{C_1}(r)$
and $t'\succ_{C_2}t$, we get a contradiction.
On the other hand, if $t'\notin\wnnw{C_1}(r)$, then by Theorem \ref{th:monotone}
we get $t'\notin\wnnw{C_2}(r)$, a contradiction.
\hfill $\Box$

Consider now what happens if the assumptions in Theorem \ref{th:commute} are relaxed.
\begin{example}
Let $Emp(EmpNo,YearEmployed,Salary)$ be a relation schema. Define the following preference
relations over it:
\[(e,y,s)\succ_{C_1}(e',y',s')\equiv s>s'\]
and 
\[(e,y,s)\succ_{C_2}(e',y',s')\equiv y<y'.\]
Clearly, neither $C_1\Rightarrow C_2$ nor $C_2\Rightarrow C_1$.
The database $r_1=\{(1,1975,100K),(2,1980,150K)\}$.
Now
\[\wnnw{C_1}(\wnnw{C_2}(r)=(1,1975,100K))\not=(2,1980,150K)=\wnnw{C_2}(\wnnw{C_1}(r)).\]
\end{example}
\begin{example}
Consider the following preference relations:
\[x\succ_{C_1}y \equiv x=a \wedge y=b\]
and 
\[x\succ_{C_2}y \equiv x=a \wedge y=b \vee x=b\wedge y=a.\]
Clearly, $C_1\Rightarrow C_2$.
However, $\succ_{C_2}$ is not a strict partial order.
We have
\[\wnnw{C_1}(\wnnw{C_2}(r))=\emptyset\not=\{a\}=\wnnw{C_2}(\wnnw{C_1}(r)).\]
\end{example}

In Theorem \ref{th:commute}, if the preference formula $C_2$ is a pure comparison ipf in $k$-DNF, then
checking the validity of the formula $\forall t_1,t_2[C_1(t_1,t_2)\Rightarrow C_2(t_1,t_2)]$
can be done in PTIME.

\subsubsection{Commuting selection and winnow}

We identify in Theorem \ref{th:selection}
below a sufficient and necessary condition under which the winnow operator
and a relational algebra selection commute.
This is helpful for pushing selections past winnow operators in 
preference queries. It is well known that moving selections down in the query
tree reduces the size of (and the time needed to materialize) intermediate results
and has a potential of enabling the use of indexes (if a selection is pushed all
the way down to a database relation that has an index matching the selection condition).

\begin{theorem}\label{th:selection}
Given a relation schema $R$, a selection condition $C_1$ over $R$ and a preference formula $C_2$ over $R$,
if the formula 
\[\forall t_1,t_2[(C_1(t_2)\wedge C_2(t_1,t_2))
\Rightarrow C_1(t_1)]\]
is valid, then for all instances $r$ of $R$:
\[\sigma_{C_1}(\wnnw{C_2}(r))=\wnnw{C_2}(\sigma_{C_1}(r)).\]
The converse holds under the assumption that $\succ_{C_2}$
is irreflexive.
\end{theorem}
{\bf Proof:}
We have that:
\[t\in\sigma_{C_1}(\wnnw{C_2}(r))\equiv t\in r\wedge C_1(t)\wedge(\neg\exists t'[t'\in r\wedge C_2(t',t)]).\]
On the other hand:
\[t\in \wnnw{C_2}(\sigma_{C_1}(r))\equiv t\in r\wedge C_1(t)\wedge
(\neg\exists t'[t'\in r\wedge C_2(t')\wedge C_2(t',t)]).\]
Clearly, the first formula implies the second.
To see that the opposite direction also holds,
assume that there is a tuple $t_0$ such that $t_0\in r$ and $C_2(t_0,t)$ holds.
 $C_1(t)$  holds, thus $C_1(t_0)$ holds too, since otherwise the formula
$\forall t_1,t_2[(C_1(t_2)\wedge C_2(t_1,t_2))
\Rightarrow C_1(t_1)]$ would not be valid.

To see the necessity of the condition of the theorem, assume that
there are tuples $t_1$ and $t_2$ such that
$C_1(t_2)\wedge C_2(t_1,t_2)\wedge \neg C_1(t_1)$.
Then
\[\wnnw{C_2}(\sigma_{C_1}(\{t_1,t_2\}))=\{t_2\}\not=\emptyset=\sigma_{C_1}(\wnnw{C_2}(\{t_1,t_2\})).\]
The irreflexivity of $\succ_{C_2}$ is necessary to ensure that
$\wnnw{C_2}(\sigma_{C_1}(\{t_1,t_2\}))$ is nonempty.
\hfill $\Box$

If the preference formula $C_2$ in Theorem \ref{th:selection} is a pure comparison ipf 
and the selection condition $C_1$ is in $k$-DNF and refers only to the arithmetic comparison predicates,
then checking the validity of the formula $\forall (C_1(t_2)\wedge C_2(t_1,t_2))\Rightarrow C_1(t_1)$
can be done in PTIME.

\begin{example}\label{ex:sal:03}
Consider the relation $Book(ISBN,Vendor,Price)$
from Example \ref{ex:sal}. 
The preference relation $\succ_{C_1}$ is defined as
\[(i,v,p)\succ_{C_1}(i',v',p') \equiv i=i' \wedge p<p'.\]
Consider the query $\sigma_{Price<15}(\wnnw{C_1}(Book))$.
Now
\[\forall p,p',i,i'[(p'<15\wedge i=i' \wedge p<p')\Rightarrow p<15]\]
is a valid formula, thus by Theorem \ref{th:selection}
\[\wnnw{C_1}(\sigma_{Price<15}(Book))=\sigma_{Price<15}(\wnnw{C_1}(Book)).\]
On the other hand, consider the query $\sigma_{Price>15}$.
Then
\[\forall p,p',i,i'[(p'>15\wedge i=i' \wedge p<p')\Rightarrow p>15]\]
is not a valid formula, thus in this case the selection does not commute
with winnow.
Finally, the query $\sigma_{ISBN=c}$ for any string $c$ commutes with
with $\wnnw{C_1}(Book)$, because
\[\forall p,p',i,i'[(i=c\wedge i=i' \wedge p<p')\Rightarrow i'=c]\]
is a valid formula.
\end{example}

\subsubsection{Commuting projection and winnow}

We deal now with projection. For winnow to commute with projection, the
preference formula needs to be restricted to the attributes in the projection.
We denote by $t[X]$  the tuple $(t[A_1],\ldots,t[A_k])$, where $X=A_1\cdots A_k$ is a set of attributes.

\begin{definition}
Given a relation schema $R$, a set of attributes $X$ of $R$, and a preference
relation $\succ_C$ over $R$, the {\em restriction $\theta_x(\succ_C)$ of $\succ_C$ to $X$} is a 
preference relation $\succ_{C'}$ defined using the following formula:
\[u\succ_{C'}u'\equiv \forall t,t'[(t[X]=u\wedge t'[X]=u')\Rightarrow t\succ_C t'].\]
\end{definition}

It is easy to see that if $\succ_C$ is a strict partial order, so is $\theta_x(\succ_C)$.

\begin{theorem}\label{th:projection}
Given a relation schema $R$, a set of attributes $X$ of $R$, and a preference formula $C$ over $R$,
if the following formulas are valid:
\[\forall t_1,t_2,t_3 [(t_1[X]=t_2[X]\wedge t_1[X]\not=t_3[X]\wedge t_1\succ_C t_3)
\Rightarrow t_2\succ_C t_3],\]
\[\forall t_1,t_3,t_4 [(t_3[X]=t_4[X]\wedge t_1[X]\not=t_3[X]\wedge  t_1\succ_C t_3)
\Rightarrow t_1\succ_C t_4 ],\]
then for all instances $r$ of $R$:
\[\pi_X(\wnnw{C}(r))=\wnnw{C'}(\pi_X(r)),\]
where $\succ_{C'}=\theta_x(\succ_C)$ is the restriction of $\succ_C$ to $X$.
The converse holds under the assumption that $\succ_C$ is irreflexive.
\end{theorem}
{\bf Proof:}
Assume $u\in\pi_X(\wnnw{C}(r))$. Then there exists a tuple $t\in\wnnw{C}(r)$ such that
$t[X]=u$. Assume $u\notin\wnnw{C'}(\pi_X(r))$. Since $u\in \pi_X(r)$,
there exists a tuple $u'\in\pi_X(r)$ such that
$u'\succ_{C'}u$ and a tuple $t'\in r$ such that $t'[X]=u'$.
Since $u'\succ_{C'}u$, it has to be the case that $t'\succ_C t$,
which contradicts the fact that $t\in\wnnw{C}(r)$.

For the opposite direction, assume that $u\in\wnnw{C'}(\pi_X(r))$
and $u\notin\pi_X(\wnnw{C}(r))$. 
Then for each tuple $t\in r$ such that $t[X]=u$, there is another tuple
$t'\in r$ such that $t'\succ_c t$ and $t'[X]\not= t[X]$.
By the assumption of the theorem, each tuple $t'$ that dominates (in $\succ_C$) one 
tuple $t$ such that $t[X]=u$, also dominates each such tuple. 
Also, any two tuples that agree on $X$ dominate the same set of tuples.
Therefore, if 
$u'=t'[X]$, then $u'\succ_{C'} u$, which contradicts the fact that $u\in\wnnw{C'}(\pi_X(r))$.

To show the converse, assume that the first condition is violated,
i.e., there are three tuples $t_1$, $t_2$ and $t_3$ such that
$t_1[X]=t_2[X]$, $t_1[X]\not=t_3[X]$, $t_1\succ_C t_3$ and $t_2\not\succ_C t_3$.
Let $r_0=\{t_1,t_2,t_3\}$.
Then $t_3\notin\wnnw{C}(r_0)$, so $\pi_X(\wnnw{C}(r_0))=\{t_1[X]\}$.
Now $t_1[X]\not\succ_{C'}t_3[X]$ (because $t_2\not\succ_C t_3$) and $t_1[X]\not=t_3[X]$.
Thus \[\wnnw{C'}(\pi_X(r))=\{t_1[X],t_3[X]\}\not=\{t_1[X]\}=\pi_X(\wnnw{C}(r_0)).\]
The violation of the second condition also leads to a contradiction in a similar way.
\hfill $\Box$

If the preference formula $C$ in Theorem \ref{th:projection} is a pure comparison ipf in $k$-DNF
then checking the validity of the assumption of this theorem
can be done in PTIME.
If $C$ is a pure-comparison ipf, then $C'$ can be presented in an equivalent, quantifier-free form.

\begin{example}\label{ex:sal:05}
Consider again the preference relation $\succ_{C_1}$ from Example \ref{ex:sal}:
\[(i,v,p)\succ_{C_1}(i',v',p') \equiv i=i' \wedge p<p'\]
over the relation schema $Book(ISBN,Vendor,Price)$.
Then the relation \mbox{$C'=\theta_{ISBN,Price}(\succ_{C_1})$} is defined as
\[(i,p)\succ_{C'}(i',p')\equiv \forall t,t'[(t[X]=(i,p)\wedge t'[X]=(i',p'))\Rightarrow t\succ_{C_1} t']
\equiv i=i'\wedge p<p'.\]
This confirms the intuition that the projection does not affect this particular
preference relation.
It is easy to see that the condition of Theorem \ref{th:projection} is also satisfied, so
winnow commutes with projection in this case.
\end{example}

\subsubsection{Distributing winnow over Cartesian product}

For winnow to distribute (in a modified form) with the Cartesian product, the preference
formula needs to be in a special form.
The form turns out to be the {\em Pareto composition}, well known in multi-attribute
utility theory \cite{Fish70}.
Preference queries involving Pareto composition are quite common:
the skyline queries \cite{BoKoSt01} without {\tt DIFF} attributes are of this form.

\begin{definition}\label{def:Pareto}
Given two relation schemas $R_1$ and $R_2$, a preference relation $\succ_{C_1}$ over $R_1$
and a preference relation $\succ_{C_2}$ over $R_2$, the {\em Pareto composition $P(\succ_{C_1},\succ_{C_2})$
of $\succ_{C_1}$ and $\succ_{C_2}$} is a preference relation $\succ_{C_0}$
over the Cartesian product $R_1\times R_2$
defined as:
\[(t_1,t_2)\succ_{C_0} (t_1',t_2')\equiv t_1\succeq_{C_1} t_1'\wedge t_2\succeq_{C_2} t_2'
\wedge (t_1\succ_{C_1} t_1'\vee t_2\succ_{C_2} t_2'),\]
where 
\[x\succeq_C y\equiv x \succ_C y \vee x=y.\]
\end{definition}

Clearly, if $\succ_{C_1}$ and $\succ_{C_2}$ are strict partial orders, so is $P(\succ_{C_1},\succ_{C_2})$.

\begin{theorem}\label{th:product}
Given two relation schemas $R_1$ and $R_2$, a preference relation $\succ_{C_1}$ over $R_1$
and a preference relation $\succ_{C_2}$ over $R_2$, for any relations $r_1$ and $r_2$
which are instances of $R_1$ and $R_2$, resp., the following property holds:
\[\wnnw{C_0}(r_1\times r_2)=\wnnw{C_1}(r_1)\times\wnnw{C_2}(r_2),\]
where $C_0=P(\succ_{C_1},\succ_{C_2})$.
\end{theorem}
{\bf Proof:}
Assume $(t_1,t_2)\in \wnnw{C_0}(r_1\times r_2)$ but 
$(t_1,t_2)\notin \wnnw{C_1}(r_1)\times\wnnw{C_2}(r_2)$.
Then $t_1\notin\wnnw{C_1}(r_1)$ or $t_2\notin\wnnw{C_2}(r_2)$.
Assume the first.
Since $(t_1,t_2)\in r_1\times t_2$ and $t_1\in r_1$,
there must be a tuple $t_1'\in r_1$ such
that $t_1'\succ_{C_1} t_1$. Then the tuple $(t_1',t_2)\in r_1\times r_2$ and
$(t_1',t_2)\succ_{C_0} (t_1,t_2)$ which contradicts the fact that $(t_1,t_2)\in \wnnw{C_0}(r_1\times r_2)$.
The second case is symmetric.

Assume now that $(t_1,t_2)\in \wnnw{C_1}(r_1)\times\wnnw{C_2}(r_2)$
and $(t_1,t_2)\notin \wnnw{C_0}(r_1\times r_2)$.
Then there is a tuple $(t_1',t_2')\in r_1\times r_2$ such that
$(t_1',t_2')\succ_{C_0}(t_1,t_2)$.
Consequently,
$t_1'\succ_{C_1} t_1$ or $t_2'\succ_{C_2} t_2$.
Both cases lead to a contradiction with the fact
that $(t_1,t_2)\in \wnnw{C_1}(r_1)\times\wnnw{C_2}(r_2)$.
\hfill $\Box$

We show now that a slight variation of the Pareto composition, even though it appears to be
more natural, fails
to achieve the distributivity of winnow over product.
\begin{example}
Define a different composition $\succ_{C_0'}$ of two preference relations $\succ_{C_1}$ and $\succ_{C_2}$
as follows:
\[(t_1,t_2)\succ_{C_0'} (t_1',t_2')\equiv t_1\succ_{C_1} t_1'\wedge t_2\succ_{C_2} t_2'.\]
Consider the following preference relations:
\[x\succ_{C_1}y\equiv x\succ_{C_2} y\equiv x>y.\]
Then if $r_1=\{1\}$ and $r_2=\{1,2\}$, then
\[\wnnw{C_1}(r_1)\times\wnnw{C_2}(r_2)=\{(1,2)\}\not=\{(1,1),(1,2)\}=\wnnw{C_0'}(r_1\times r_2).\]
\end{example}

\subsubsection{Distributing winnow over union and difference}

It is possible to distribute winnow over union or difference only in the trivial
case where the preference relation is an anti-chain.
We call two relation schemas {\em compatible} if they have the same number of attributes
and the corresponding attributes have the same domains.
\begin{theorem}\label{th:union}
Given two compatible relation schemas $R$ and $S$ and an irreflexive preference relation $\succ_C$ over $R$,
we have for every relation $r$ and $s$
\[\wnnw{C}(r\cup s)=\wnnw{C}(r)\cup\wnnw{C}(s)\]
and
\[\wnnw{C}(r- s)=\wnnw{C}(r)-\wnnw{C}(s)\]
if and only if $\succ_C=\emptyset$.
\end{theorem}
{\bf Proof:}
Clearly, if $\succ_C=\emptyset$ and $\succ_C$ is irreflexive,  then
\[\wnnw{C}(r)\cup\wnnw{C}(s)=r\cup s=\wnnw{C}(r\cup s).\]
To show that this is a necessary condition, assume that $\succ_C\not=\emptyset$.
Then there are two tuples $t_1$ and $t_2$ such that $t_1\succ_C t_2$.
Now
\[\wnnw{C}(\{t_1,t_2\})=\{t_1\}\not=\{t_1,t_2\}=\wnnw{C}(\{t_1\})\cup\wnnw{C}(\{t_2\}).\]
The proof for difference is similar.
\hfill $\Box$

\section{Composition of preferences}\label{sec:spec}

Preference relations may be composed in many different ways.
In general, we distinguish between {\em multi-dimensional} and {\em uni-dimensional}
composition.
In multi-dimensional composition, we have a number of preference relations
defined over several database relation schemas, and we define a preference relation
over the Cartesian product of those relations. 
An example is Pareto composition (Definition \ref{def:Pareto}).
Another example is lexicographic composition.
In uni-dimensional composition, a number of preference relations over a single
database schema are composed, producing another preference relation over the same schema.
Examples include: Boolean and prioritized composition (discussed below).

Since in our framework preference relations are defined by first-order preference formulas,
any first-order definable composition
of preference relations leads again to first-order
preference formulas, which in turn can be used as parameters of the winnow
operator.
The composition does not even have to be first-order definable, as long as
it produces a (first-order) preference formula. We'll see an example of the latter
later in section when we discuss transitive closure.

\subsection{Boolean composition}

Union, intersection and difference of preference relations
are obviously captured by the Boolean operations on the
corresponding preference formulas. 
For example, the following formula captures the preference $\succ_{C_0}=\succ_{C_1}\cap\succ_{C_2}$:
\[x\succ_{C_0} y\equiv x\succ_{C_1} y \wedge x\succ_{C_2}y.\]

Table \ref{tab:boolean}
summarizes the preservation of  properties of relations
by the appropriate Boolean composition operator.
\begin{table}[htb]
\begin{center}
\begin{tabular}{|l|l|l|l|}\hline
& Union &Intersection &Difference\\\hline
Irreflexivity & Yes & Yes &Yes\\\hline
Asymmetry& No &Yes &Yes\\\hline
Transitivity &No & Yes &No\\\hline
\end{tabular}
\end{center}
\caption{Properties Preserved by Boolean Composition}
\label{tab:boolean}
\end{table}

\subsection{Preference hierarchies}
It is often the case that preferences form hierarchies.
For instance, I may have a general preference for red wine
but in specific cases, e.g., when eating fish, this preference is overridden
by the one for white wine.
Also a preference for less expensive books (Example \ref{ex:sal}) can be
overridden by a preference for certain vendors.

\begin{definition}\label{def:priority}
Consider two preference relations $\succ_{C_1}$ and $\succ_{C_2}$
defined over the same schema $U$.
The {\em prioritized composition} $\succ_{C_{1,2}}=\succ_{C_1} \rhd \succ_{C_2}$ of 
$\succ_{C_1}$ and $\succ_{C_2}$ is defined as:
\[t_1\succ_{C_{1,2}}t_2\equiv t_1\succ_{C_1}t_2 \vee
(t_2\not\succ_{C_1}t_1\wedge t_1\succ_{C_2}t_2).\]
\end{definition}

The prioritized composition $\succ_{C_1} \rhd \succ_{C_2}$ has the following
intuitive reading: {\em prefer according to $\succ_{C_2}$ unless $\succ_{C_1}$
is applicable.}

\begin{example}\label{ex:sal:1}
Continuing Example \ref{ex:sal},
instead of the preference relation $\succ_{C_1}$ defined there as follows:
\[(i,v,p)\succ_{C_1}(i',v',p') \equiv i=i' \wedge p<p',\]
we consider
the relation $\succ_{C_0}\rhd\succ_{C_1}$ where $\succ_{C_0}$ is defined
by the following formula $C_0$:
\[(i,v,p)\succ_{C_0}(i',v',p') \equiv i=i' \wedge v={\rm 'BooksForLess'} \wedge
v'={\rm 'LowestPrices'}.\]
Assume the preference relation $\succ_{C_{0,1}}=\succ_{C_0}\rhd\succ_{C_1}$ 
(the definition of $\succ_{C_{0,1}}$ is easily obtained from the formulas $C_0$ and $C_1$
by substitution).
Then $\wnnw{C_{0,1}}(r_1)$ returns the following tuples
\begin{center}
\begin{tabular}{|l|l|l|}
\hline
{\em ISBN} &{\em Vendor} &{\em Price}\\\hline
0679726691 & BooksForLess & \$14.75\\
0062059041 & BooksForLess & \$7.30\\
0374164770 & LowestPrices & \$21.88\\\hline
\end{tabular}
\end{center}
Note that now a more expensive copy of the first book is preferred, due
to the preference for 'BooksForLess' over 'LowestPrices'.
However, 'BooksForLess' does not offer the last book, and that's
why the copy offered by 'LowestPrices' is preferred.
\end{example}

\begin{theorem}\label{th:fod}
If $\succ_{C_1}$ and $\succ_{C_2}$ are preference relations,
so is $\succ_{C_{1,2}}$. 
If $\succ_{C_1}$ and $\succ_{C_2}$ are both irreflexive or asymmetric, so is $\succ_{C_{1,2}}$.
\end{theorem}

However, a relation defined as the prioritized composition of two transitive preference
relations does not have to be transitive.
\begin{example}\label{ex:not:trans}
Consider the following preference relations:
\[a\succ_{C_1} b, b\succ_{C_2} c.\]
Both $\succ_{C_1}$ and $\succ_{C_2}$ are trivially transitive.
However, $\succ_{C_1} \rhd \succ_{C_2}$ is not.
\end{example}

\begin{theorem}\label{th:assoc}
Prioritized composition is associative:
\[(\succ_{C_1}\rhd\succ_{C_2})\rhd\succ_{C_3}\equiv\succ_{C_1}\rhd(\succ_{C_2}\rhd\succ{C_3})\]
and distributes over union:
\[\succ_{C_1}\rhd(\succ_{C_2}\cup\succ_{C_3})\equiv(\succ_{C_1}\rhd\succ_{C_2})\cup
(\succ_{C_1}\rhd\succ_{C_3}).\]
\end{theorem}

Thanks to the associativity and distributivity of $\rhd$,
the above construction can  be generalized to an arbitrary finite partial {\em priority order}
between preference relations. Such an order can be viewed as a graph in which
the nodes consist of preference relations and the edges represent relative
priorities (there would be an edge $(\succ_{C_1},\succ_{C_2})$ in the situation
described above). To encode this graph as a single preference relation,
one would construct first the definitions corresponding to individual
paths from roots to leaves, and then take a disjunction of all such
definitions.

\ignore{
This approach can also be applied to model defeasible
and default preferences.}

There are many other ways of combining preferences. 
For instance, the paper \cite{AnRySc02} defines an infinite family of uni-dimensional composition
operators for preference relations on the basis of two basic operators. 
Since all the definitions are first-order, every preference relation defined in the
framework of  \cite{AnRySc02} can also be defined in ours.
In \cite{AnRySc02}, it is proved that the operators in the defined
family exhaust all operators  satisfying a number of intuitively plausible postulates.
It turns out that the operator $\rhd$ defined above cannot be captured in the
framework of \cite{AnRySc02}, because it violates one of those postulates: it does not preserve transitivity.

\subsection{Transitive closure}

We address here the issue of {\em transitively closing} a preference relation.
We have seen an example (Example \ref{ex:sal}) 
of a preference relation that is already transitive.
However, there are cases when we expect the preference
relation to be the {\em transitive closure} of another preference relation
which is not transitive.

\begin{example}\label{ex:trans}
Consider the following relation:
\[x\succ y\equiv x=a \wedge y=b \vee x=b\wedge y=c.\]
In this relation, $a$ and $c$ are not related though
there are contexts in which this might be natural.
(Assume I prefer to walk than to drive, and to drive than
to ride a bus. Thus, I also prefer to walk than to ride a bus.)
\end{example}

In our framework, we can specify the preference
relation $\succ_{C^*}$ to be the {\em transitive closure} of
another preference relation $\succ$ defined using a first-order formula.
This is similar to transitive closure queries in relational
databases. However, there is an important difference.
In databases, we are computing the transitive closure of
a {\em finite} relation, while here we are transitively closing
an infinite relation defined using a first-order formula.

\begin{definition}\label{def:transitive}
The {\em transitive closure} of a  preference relation
$\succ_C$ over a relation schema $R$ is a  preference relation 
$\succ_{C^*}$ over $R$ defined as: 
\[t_1\succ_{C^*}t_2\;{\rm iff}\; t_1\succ_C^nt_2\;
{\rm for\; some\;} n\geq 0,\]
where:
\[\begin{array}{l}
t_1\succ_C^1 t_2\equiv t_1\succ_C t_2\\
t_1\succ_C^{n+1}t_2\equiv\exists t_3.\ 
t_1\succ_C t_3\wedge t_3\succ_C^n t_2.\\
\end{array}\]
\end{definition}

Clearly, in general Definition \ref{def:transitive} leads to infinite formulas.
However, as Theorem \ref{th:trans} shows, in many important cases the preference
relation $\succ_{C*}$ will in fact be defined by a finite formula.

\ignore{
\begin{theorem}\label{th:transitive}
Let $\succ_C$ be a  preference relation over a relation schema $R$.
Then for every instance $r$ of $R$:
\[\wnnw{C}(r)=\wnnw{C^*}(r).\]
\end{theorem}
{\bf Proof:}
Since $C\Rightarrow C*$, then $\wnnw{C^*}(r)\subseteq \wnnw{C}(r)$.
To establish the containment in the opposite direction,
assume there is a tuple $t\in \wnnw{C}(r)$ but $t\notin\wnnw{C^*}(r)$.
Then there is a tuple $t'\in \wnnw{C^*}(r)$ such that
$t'\succ_{C^*} t$. But then according to the definition
of $\succ_{C^*}, there is a tuple $t''
}

\begin{theorem}\label{th:trans}
If a preference relation $\succ_C$ is defined using a pure comparison ipf,
the transitive closure $\succ_{C^*}$
of $\succ_C$ is also defined using a pure comparison ipf
and that definition can be effectively obtained.
\end{theorem}
{\bf Proof:} 
The computation of the transitive closure can in this case be
formulated as the evaluation of Datalog with order or gap-order
(for integers) constraints.
Suppose $\succ_C$ is defined as:
\[x\succ_C y \equiv \alpha_1(x,y)\vee\cdots\vee\alpha_n (x,y).\]
Then the Datalog program that computes the formula $C^*$ defining $\succ_{C^*}$
looks as follows:
\[\begin{array}{l}
T(x,y) \leftarrow \alpha_1(x,y).\\
\cdots\\
T(x,y) \leftarrow \alpha_n(x,y).\\
S(x,y)\leftarrow T(x,y).\\
S(x,y)\leftarrow T(x,z), S(z,y).\\
\end{array}\]
The evaluation of this program terminates \cite{KaKuRe95,RevTCS93} and its result,
collected in $S$, represents the desired formula.

\ignore{
In this case the existential quantifiers can be removed
using quantifier elimination.
To obtain the required first-order definition of the transitive
closure, we have to compare for equivalence 
$\bigvee_{k\geq n\geq 1}\Phi^n(t_1,t_2)$ 
and $\bigvee_{k+1\geq n\geq 1}\Phi^n(t_1,t_2)$, for consecutive
values of $k$.
Since both formulas are quantifier-free, such a comparison
is decidable for the given signatures.}
$\hfill\Box$

An analogous result holds if instead of arithmetic comparisons we
consider equality constraints over an infinite domain \cite{KaKuRe95}.
\ignore{
Also, similar techniques can be used to provide a PTIME decision
procedure for acyclicity in the cases listed above.}

\begin{example}\label{ex:trans:1}
Continuing Example \ref{ex:trans}, we obtain the following
preference relation $\succ_{C^*}$ by transitively closing $\succ_C$:
\[x\succ_{C^*} y\equiv x=a \wedge y=b \vee x=b\wedge y=c \vee x=a\wedge y=c.\]
\end{example}

Theorem \ref{th:trans} is not in conflict with the well-known non-first
order definability of transitive closure on finite structures.
In the latter case it is shown that there is no finite first-order formula
expressing
transitive closure for arbitrary (finite) binary relations. In Theorem 
\ref{th:trans} the relation to be closed, although possibly infinite,
is fixed (since it is defined using the given ipf). In particular, given an encoding
of a fixed finite binary relation using an ipf, the transitive closure
of this relation is defined using another ipf.

The transitive closure of a irreflexive (resp. asymmetric)
preference relation may fail to be irreflexive (resp. asymmetric).

\section{Applications and extensions}\label{sec:ext}

We show here how to use winnow to express special classes of
preference queries: skylines and queries involving scoring functions, and how to use
winnow together with other operators of the relational algebra to
express more complex decision problems involving preferences. We
consider the following: integrity constraints, extrinsic preferences,
and aggregation.  

\subsection{Special classes of preference queries}

\subsubsection{Skylines}
{\em Skyline queries} \cite{BoKoSt01} find all the tuples in a relation
that {\em are not dominated by any other tuples in the same relation in all dimensions}.
This is exactly the notion of Pareto composition (Definition \ref{def:Pareto}) in an arbitrary
number of dimensions.

\begin{figure}[htb]
\centering
\psset{xunit=1cm}
\psset{yunit=1cm}
\begin{pspicture}(0,0)(6,3)
\psaxes[ticks=none,Dx=1,Dy=1]{->}(6,3)
\qdisk(1,0.5){1pt}
\qdisk(2,1){1pt}
\qdisk(2,2.25){3pt}
\qdisk(3,1.75){1pt}
\qdisk(3.5,1.5){1pt}
\qdisk(4,2){3pt}
\qdisk(4.5,0.75){1pt}
\qdisk(5,1){3pt}
\end{pspicture}
\caption{Two-dimensional skyline}
\label{fig:skyline}
\end{figure}
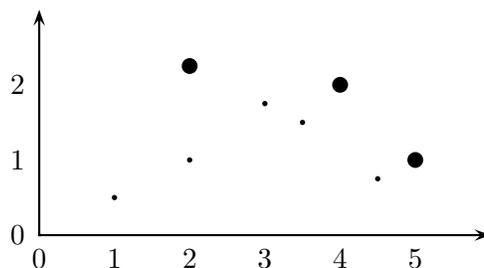

Figure \ref{fig:skyline} shows an example of a two-dimensional skyline where
the dominance relationship is $>$.
The skyline elements are marked with thick black dots.

\cite{BoKoSt01} propose to write skyline queries using the following extension to SQL:
\begin{verbatim}
  SELECT ... FROM ... WHERE ...
  GROUP BY ... HAVING ...
  SKYLINE OF A1 [MIN | MAX | DIFF]
             ...
             An [MIN | MAX | DIFF]
\end{verbatim}

The values of a {\tt MIN} attribute are minimized, those of a {\tt MAX} attribute maximized.
A {\tt DIFF} attribute indicates that tuples with {\em different} values of that attribute
are incomparable. 
The {\tt SKYLINE} clause is applicable after all other SQL clauses.

Clearly, skylines can be expressed using winnow. The winnow is applied to an SQL view
that expresses the non-skyline constructs in a skyline query.
The preference formula is easily obtained from the {\tt SKYLINE} clause.
For example:
\begin{verbatim}
SKYLINE OF A DIFF, B MAX, C MIN
\end{verbatim}
in a relation $R$ is equivalent to $\wnnw{C}(R)$ where
\[(x,y,z)\succ_C(x',y',z')\equiv x=x'\wedge y\geq y' \wedge z\leq z' \wedge (y>y'\vee z<z').\]
\vspace{.2in}

Finally, we note that $\wnnw{C_1}(Book)$ from Example \ref{ex:sal:02} is also a skyline
query in which the skyline clause looks as follows:
\begin{verbatim}
SKYLINE OF ISBN DIFF, PRICE MIN.
\end{verbatim}

\subsubsection{Queries involving scoring functions}

Sometimes a relation schema $R$ comes with a scoring function that associates
a value $f(t)$ with every tuple $t$ in a possible instance of $R$.
Now finding the tuples that maximize a scoring function $f$ can be done
by computing $\wnnw{C_f}(r)$ for the given instance $r$ of $R$, where
\[t\succ_{C_f}t'\equiv f(t)>f(t').\]

This approach can be generalized to compute not only the top scoring tuples
but also those whose score differs from the top score by at most a given value or
a given percentage. For example, the tuples that differ from the top score by at most $d$
are computed by $\wnnw{C_{f-d}}(r)$, where
\[t\succ_{C_{f-d}}t'\equiv f(t)-d>f(t').\]

Queries that return the tuples with {\em top-N} scores \cite{CaKo97}
can also be captured using winnow together with SQL, using the approach described
later in this section. Essentially, for each tuple $t$ we will determine using SQL the number $n(t)$ of tuples
with higher scores than $t$ and use the expression $N-n(t)$, where is the number of tuples in the relation,
to define a new scoring function. This function is then used to define a preference relation
as in the preceding paragraph.
It appears, however, that in terms of the efficiency of query evaluation this approach will be inferior 
to the approach in which top-N queries are supported directly by the query engine.

Formally,we say that a real-valued function $f$ over a schema $R$ 
{\em represents} a preference relation $\succ_C$ over $R$ iff
\[\forall t_1,t_2\ [t_1\succ_C t_2 \;{\rm iff}\; f(t_1)>f(t_2).\]
As pointed out earlier, not every preference relation which is a strict partial order can be expressed
using a scoring function. A necessary condition is that the relation be a {\em weak order} \cite{Fish70}. 
We can ask for the motivation behind this notion of representation.
It is easy to show that
\begin{theorem}
A real-valued function $f$ represents a preference relation $\succ_C$ iff
for every finite instance $r$ of $R$, the set $\wnnw{C}(r)$ is equal to the set
of tuples of $r$ assuming the maximum value of $u$.
\end{theorem}
Thus, if a scoring function does not represent a preference relation, that fact can be detected
by winnow evaluated over some instance.

There are other, weaker forms of representation of preference relations by scoring functions.
For instance, if we only require that
\[\forall t_1,t_2\ [t_1\succ_C t_2 \Rightarrow f(t_1)>f(t_2),\]
then for every strict partial order there is a scoring function representing it.
However, in this case we can only guarantee that the set of tuples in a given instance $r$ that
maximize $f$ is a {\em subset} of $\wnnw{C}(r)$.

\subsection{Integrity constraints}

There are cases when we wish to impose a constraint on the
result of the winnow operator. In Example \ref{ex:sal},
we may say that we are interested only in the books under \$15.
In Example \ref{ex:wine}, we may restrict our attention only
to the meat or fish dishes (note that currently the dishes that
are not meat or fish do not have a preferred kind of wine).
In the same example, we may ask for a specific number of meal recommendations.

In general, we need to distinguish between {\em local} and
{\em global} constraints. A local constraint imposes a condition
on the components of a single tuple,
for instance {\tt Book.Price<\$15}.
A global constraint imposes
a condition on a set of tuples. The first two examples above are local constraints; the
third is global. To satisfy a global constraint on the result of the winnow operator,
one would have to construct a maximal subset of this answer
that satisfies the constraint. Since in general there may be more
than one such subset, the required construction cannot be described
using a single relational algebra query.
On the other hand, local constraints are easily handled, 
since they can be expressed using selection.
In general, it matters whether the selection is applied before or after the winnow operator.
Theorem \ref{th:selection} identifies sufficient and necessary conditions for winnow
and selection to commute.

\begin{example}\label{ex:cars}
Consider the situation where we have a specific preference ordering for cars, e.g.,
prefer BMW to Chevrolet, but also have a limited budget (captured by a selection condition).
Then clearly, selecting the most desirable affordable car will not give the same result
as selecting the most desirable cars if they are affordable.
\end{example}

A {\em veto} expresses a prohibition on the presence of a specific set of values in 
the elements of the answer to a preference query and thus can be viewed as
a local constraint.
To veto a specific tuple $w=(a_1,\ldots,a_n)$ in a relation $S$ (which can be defined
by a preference query) of arity $n$, we 
write the selection:
\[\sigma_{A_1\not=a_1\vee\cdots\vee A_n\not=a_n}(S).\]
\ignore{
We can also veto a tuple directly in the result of the winnow
operator $\wnnw{C}(R)$ by modifying the preference formula
to $C\vee C_w$ where $C_w$ is defined as 
\[(x_1,\ldots,x_n)\succ_w (y_1,\ldots,y_n)\equiv
x_1=a_1\wedge\cdots\wedge x_n=a_n\wedge y_1=a_1\wedge\cdots\wedge y_n=a_n.\]
}
\subsection{Intrinsic vs. extrinsic preferences}

So far we have talked only about {\em intrinsic} preference formulas.
Such formulas establish the preference relation between two tuples purely
on the basis of the values occurring in those tuples.
{\em Extrinsic} preference formulas may refer not only to built-in predicates but
also to other constructs, e.g., database relations.
In general, extrinsic preferences
can use a variety of criteria: properties of the relations
from which the tuples were selected, properties of other relations, 
or comparisons of aggregate values, and do not even have to be defined 
using first-order formulas.

It is possible to express some extrinsic preferences using the winnow
operator together with other relational algebra operators using
the following multi-step strategy:
\begin{enumerate}
\item using a relational query, combine all the information relevant for the preference in a single
relation,
\item apply the appropriate winnow operator to this relation,
\item project out the extra columns introduced in the first step.
\end{enumerate}

The following example demonstrates the above strategy, as well as the use of aggregation
for the formulation of preferences.

\begin{example}\label{ex:sal:6}
Consider again the relation $Book(ISBN,Vendor,Price)$.
Suppose for each book a preferred vendor (there may be more than one)
is a vendor that sells the {\em maximum
total} number of books. 
Clearly, this is an extrinsic preference since it cannot be established 
solely by comparing pairs of tuples from this relation.
However, we can provide the required aggregate values and connect them
with individual books through new, separate views:
\begin{verbatim}
CREATE VIEW BookNum(Vendor,Num) AS
  SELECT B1.Vendor, COUNT(DISTINCT B1.ISBN)
  FROM Book B1
  GROUP BY B1.Vendor;

CREATE VIEW ExtBook(ISBN,Vendor,Num) AS
  SELECT B1.ISBN, B1.Vendor, BN.Num
  FROM Book B1, BookNum BN
  WHERE B1.Vendor=BN.Vendor;
\end{verbatim}

Now the extrinsic preference is captured by the query 
\[\pi_{ISBN,Vendor}(\wnnw{C_5}(ExtBook))\]
where the preference formula $C_5$ is defined as follows:
\[(i,v,n)\succ_5 (i',v',n') \equiv i=i' \wedge n>n'.\]
\end{example}

\begin{example}
To see another example of extrinsic preference, consider
the situation in which we prefer any tuple from a relation
$R$ over any tuple from a relation $S$ which is disjoint from $R$.
Notice that this is truly an extrinsic preference, since
it is based on where the tuples come from and not on their
values. It can be handled in our approach by {\em tagging} the
tuples with the appropriate relation names (easily done in relational
algebra or SQL) and then defining the preference relation
using the tags.
If there is a tuple which belongs both to $R$ and $S$, then the
above preference relation will fail to be irreflexive and the
simulation using intrinsic preferences will not work.
Note also that an approach similar to tagging was used in Example \ref{ex:wine}
(wine and dish types play the role of tags).
\end{example}
\begin{example}
Suppose user preferences are stored in a database relation {\it Pref(A,B)}.
Then once can define an extrinsic preference relation:
\[x\succ_{Pref} y \equiv Pref(x,y).\]
Such a preference relation cannot be defined using a pure comparison ipf,
because the transitive closure of a preference relation defined using an ipf
is finite (Theorem \ref{th:trans}), while that of $\succ_{Pref}$ is infinite.
\end{example}

\section{Iterated preferences and ranking}\label{sec:iter}

We show here that the framework presented so far can be further developed
to capture other preference-related concepts like ranking. We also
present a variant of winnow suitable to preference relations that are not
partial orders.

\subsection{Ranking}
A natural notion of {\em ranking} is implicit in our approach.
A ranking is defined using {\em iterated preference}.

\begin{definition}\label{def:iterwinnow}
Given a preference relation $\succ$ defined by a pf $C$,
the {\em n-th iteration} of the winnow operator $\wnnw{C}$ in $r$ is defined as:
\[\begin{array}{l}
\wnnw{C}^1(r)=\wnnw{C}(r)\\
\wnnw{C}^{n+1}(r)=\wnnw{C}(r-\bigcup_{1\leq i\leq n}\wnnw{C}^i(r))\\
\end{array}\]
\end{definition}

For example, the query $\wnnw{C}^2(r)$ computes the set of
``second-best'' tuples.
\begin{example}\label{ex:sal:3}
Continuing Example \ref{ex:sal}, the query $\wnnw{C_1}^2(r_1)$
returns
\begin{center}
\begin{tabular}{|l|l|l|}
\hline
{\em ISBN} &{\em Vendor} &{\em Price}\\\hline
0679726691 & BooksForLess & \$14.75\\\hline
\end{tabular}
\end{center}
and the query $\wnnw{C_1}^3(r_1)$ returns
\begin{center}
\begin{tabular}{|l|l|l|}
\hline
{\em ISBN} &{\em Vendor} &{\em Price}\\\hline
0679726691 & QualityBooks & \$18.80\\\hline
\end{tabular}
\end{center}

\end{example}
Therefore, by iterating the winnow operator one can {\em rank} the
tuples in a given relation instance.

\begin{theorem}\label{th:ranking}
If a preference relation $\succ_C$ over a relation schema $R$
is a strict partial order, then for every finite instance $r$ of $R$ and every tuple $t\in r$,
there exists an $i$, $i\geq 1$, such that $t\in\wnnw{C}^i(r)$.
\end{theorem}
{\bf Proof:}
Assume there is a tuple $t_0\in \wnnw{C}(r)$ such that for all $i\geq 1$, 
$t_0\notin\wnnw{C}^i(r)$.
Select the least $i_0$ such that $\forall i\geq i_o$, $\wnnw{C}^i(r)=\emptyset$
(such an $i_0$ always exists due to the finiteness of $r$).
Clearly, $t_0\notin\wnnw{C}^{i_0}(r)$, thus $t\in r-\bigcup_{1\leq i\leq {i_0-1}}\wnnw{C}^i(r)$.
Then there must be a tuple $t_1$ such that $t_1\succ_C t_0$ and
$t_1\in r-\bigcup_{1\leq i\leq {i_0-1}}\wnnw{C}^i(r)$ (otherwise $t_0\in\wnnw{C}^{i_0}(r)$).
Since $\succ_C$ is a strict partial order, there has to be an infinite increasing chain
in $r$, a contradiction with the finiteness of $r$.
$\hfill\Box$

We define now the ranking operator $\rnk{C}(R)$.
\begin{definition}
If $R$ is a relation schema and $C$ a preference
formula defining a preference relation $\succ_C$ over $R$,
then the {\em ranking operator} is written as $\rnk{C}(R)$, 
and for every instance $r$ of $R$:
\[\rnk{C}(r)=\{(t,i)\mid t\in \wnnw{C}^i(r)\}.\]
\end{definition}

One can now study the algebraic properties of the ranking operator, that parallel
those that we established for winnow in Section \ref{sec:winnow}.
We list here only one property which is the most important one from a practical point of view:
commutativity of selection with ranking. In this context, ranking enjoys identical
properties to winnow.
\begin{theorem}\label{th:selection:ranking}
Given a relation schema $R$, a selection condition $C_1$ over $R$ and a preference formula $C_2$ over $R$,
if the formula 
\[\forall t_1,t_2[(C_1(t_2)\wedge C_2(t_1,t_2))
\Rightarrow C_1(t_1)]\]
is valid, then for all instances $r$ of $R$:
\[\sigma_{C_1}(\rnk{C_2}(r))=\rnk{C_2}(\sigma_{C_1}(r)).\]
The converse holds under the assumption that $\succ_{C_2}$
is irreflexive.
\end{theorem}
{\bf Proof:}
The proof is by induction on tuple rank.
The base case follows from Theorem \ref{th:selection}
and the inductive case from the observation that
\[\sigma_{C_1}(\wnnw{C_2}^{n+1}(r))=\sigma_{C_1}(\wnnw{C_2}(r-\bigcup_{1\leq i\leq n}\wnnw{C}^i(r)))=
\wnnw{C_2}(\sigma_{C_1}(r-\bigcup_{1\leq i\leq n}\wnnw{C}^i(r)))\]
which is equal to
\[\wnnw{C_2}(\sigma_{C_1}(r)-\sigma_{C_1}(\bigcup_{1\leq i\leq n}\wnnw{C}^i(r)))=
\wnnw{C_2}(\sigma_{C_1}(r)-\bigcup_{1\leq i\leq n}\sigma_{C_1}(\wnnw{C}^i(r))).\]
under the assumptions of the theorem.
$\hfill\Box$
\ignore{
It would be interesting to consider the evaluation of queries
of the form
\[\bigcup_{1\leq n\leq k}\wnnw{C}^n(r).\]
Also, are infinitary queries 
\[\bigcup_{n\geq 1}\wnnw{C}^n(r)\]
first-order definable? Probably not but then in what language to
formulate them?
}
\subsection{Weak winnow}
If a preference relation is not a strict partial order, then Theorems
\ref{th:nonempty} and  \ref{th:ranking}
may fail to hold. A number of tuples can block each other from appearing
in the result of any iteration of the winnow operator.
However, even in this case there may be a weaker form of ranking available.
\begin{example}\label{ex:sal:4}
Consider Examples \ref{ex:sal} and \ref{ex:sal:1}.
If the preference formula $C'$ is defined as $C_0\vee C_1$,
then the first two tuples of the instance $r_1$ block each other
from appearing in the result of $\wnnw{C'}(r_1)$, since 
according to $C_0$ the first tuple is preferred to the second
but just the opposite is true according to $C_1$.
Intuitively, both those tuples should be preferred to
(and ranked higher) than the third tuple.
But since neither the first not the second tuple is a member of $\wnnw{C'}(r_1)$,
none of the first three tuples can be ranked.
\end{example}

To deal with preference relations that are not strict partial orders, 
we define a new, weaker form of the winnow operator.
We relax the asymmetry and irreflexivity requirements but preserve transitivity.

To define this operator, we notice that as long as the preference relation $\succ_C$
is transitive, we can use it to define another preference relation $\succ_{C_>}$ which is a
strict partial order:
\[x\succ_{C_>}y \equiv x\succ_C y \wedge y\not\succ_C x.\]

\begin{definition}\label{def:weakwinnow}
If $R$ is a relation schema and $\succ_C$ a transitive preference relation over $R$,
then the {\em weak winnow operator} is written as $\wwnnw{C}(R)$ 
and for every instance $r$ of $R$, $\wwnnw{C}(r)=\wnnw{C_>}(r)$.
\end{definition}
 
It follows from the definition that 
\[\wwnnw{C}(r)=\{t\in r \mid \forall t'\in r.\ 
t\succ_C t'\vee t'\not\succ_C t\}.\] 
Thus the weak winnow operator returns all
the tuples that are dominated only by the tuples that they dominate
themselves.

\begin{example}\label{ex:sal:5}
Considering Example \ref{ex:sal:4}, we see that
the query $\wwnnw{C'}(r_1)$ returns now 
\begin{center}
\begin{tabular}{|l|l|l|}
\hline
{\em ISBN} &{\em Vendor} &{\em Price}\\\hline
0679726691 & BooksForLess & \$14.75\\
0679726691 & LowestPrices & \$13.50\\
0062059041 & BooksForLess & \$7.30\\
0374164770 & LowestPrices & \$21.88\\\hline
\end{tabular}
\end{center}
\end{example}

Below we formulate a few properties of the weak winnow operator.
Using Theorems \ref{th:monotone} and \ref{th:nonempty} (notice that $C_>\Rightarrow C$),
we immediately obtain the following theorem.
\begin{theorem}
If $R$ is a relation schema and $\succ_C$ a transitive preference relation over $R$,
then:
\begin{itemize}
\item for every instance $r$ of $R$, $\wnnw{C}(r)\subseteq\wwnnw{C}(r)$.
\item for every finite, nonempty relation instance $r$ of $R$, $\wwnnw{C}(r)$
is nonempty.
\end{itemize}
\end{theorem}

One can define the iteration of the weak winnow operator similarly to that
of the winnow operator (Definition \ref{def:iterwinnow}).

\begin{theorem}\label{th:weakranking}
If a preference relation $\succ_C$ over a relation schema $R$ 
is transitive, then for every finite instance  $r$ of $R$ and for every tuple $t\in r$,
there exists an $i$, $i\geq 1$, such that $t\in\wwnnw{C}^i(r)$.
\end{theorem}

\section{Related work}\label{sec:related}
\subsection{Preference queries}
\cite{LaLa87} originated the study of {\em preference queries}.
It proposed an extension of the relational calculus in which preferences
for tuples satisfying given logical conditions can be expressed.
For instance, one could say: {\em Among the tuples of $R$ satisfying $Q$, I prefer
those satisfying $P_1$; among the latter I prefer those satisfying $P_2$}.
Such a specification was to mean the following:
Pick the tuples satisfying $Q\wedge P_1\wedge P_2$; if the result is empty,
pick the tuples satisfying $Q\wedge P_1\wedge \neg P_2$; if the result is empty,
pick the remaining tuples of $R$ satisfying $Q$.
This can be simulated in our framework
as the relational algebra expression $\wnnw{{C^*}}(\sigma_Q(R))$
where ${C^*}$ is an ipf defined in the following way:
\begin{enumerate}
\item obtain the formula $C$ defining a preference relation $\succ$
\[t_1\succ t_2 \equiv P_1(t_1)\wedge P_2(t_1) \wedge
P_1(t_2)\wedge \neg P_2(t_2) \vee
P_1(t_1)\wedge \neg P_2(t_1)\wedge \neg P_1(t_2),\]
\item transform $C$ into DNF to obtain an ipf $C'$, and 
\item close the result
transitively to obtain an ipf ${C^*}$ defining
a transitive preference relation $\succ^*$
(as described in Section \ref{sec:spec}). 
\end{enumerate}
Other kinds of logical conditions from \cite{LaLa87} can be similarly
expressed in our framework. 
Maximum/minimum value preferences (as in Example \ref{ex:sal}) are
handled in \cite{LaLa87} through the explicit use of aggregate functions.
The use of such functions is implicit in the definition of our
winnow operator.

Unfortunately, \cite{LaLa87} does not contain a 
formal definition of the proposed language, so a complete
comparison with our approach is not possible. It should be noted,
however, that the framework of \cite{LaLa87} seems unable to capture
very simple conditional preferences like the ones in Examples
\ref{ex:wine} and \ref{ex:trans}. Also, it can only handle 
strict partial orders of bounded depth (except in the case
where aggregate functions can be used, as in Example \ref{ex:sal}).
Hierarchical or iterated preferences are not considered.

\cite{GoJaMa01} was
one of the sources of inspiration for the present paper.  It defines
{\em Preference Datalog}: a combination of Datalog and
clausally-defined preference relations. Preference Datalog captures,
among others, the class of preference queries discussed in
\cite{LaLa87}.  The declarative semantics of Preference Datalog is based on the
notion of {\em preferential consequence}, introduced earlier by the
authors in \cite{GoJaMa95}.  This semantics requires preferences to be
reflexive and transitive. Also, the operational semantics of
Preference Datalog uses specialized versions of the standard 
logic program evaluation methods: bottom-up \cite{GoJaMa01}
or top-down \cite{GoJaMa95}. In the context of database queries,
the approach proposed in the present paper achieves similar goals
to that of \cite{GoJaMa95} and \cite{GoJaMa01}, remaining, however,
entirely within the relational data model and classical first-order logic.
Finally, \cite{GoJaMa95,GoJaMa01} do not address some of the issues 
we deal with in the present paper like transitive closure of
preferences, prioritized composition or iterated preferences
(a similar concept to the last one is presented under the name
of ``relaxation''). More importantly, the issues of embedding the framework
into a real relational query language and optimizing preference
queries are not addressed.

\cite{Kie02,KiKo02} propose  an (independently developed) framework similar to the one presented
in this paper and in \cite{ChEDBT02}. 
A formal language for formulating preference relations is described. The language
has a number of base preference constructors and their combinators 
(Pareto and lexicographic composition, intersection, disjoint union and others).
Clearly, all of those can be captured in our framework.
On the other hand, \cite{Kie02,KiKo02} do not consider the possibility of having
arbitrary operation and predicate signatures in preference formulas, and do not
identify any specific classes of preference formulas. Neither do they consider
extrinsic preferences, complex preferences involving aggregation, or ranking.
However, the embedding into relational query languages they use is identical
to ours (it is called Best Match Only, instead of winnow). While some possible
rewritings for preference queries are presented in \cite{Kie02}, abstract properties
of winnow that we described in Section \ref{sec:winnow} are not identified.
Finally, \cite{KiKo02} describes an implementation of the framework of \cite{Kie02}
using a language called Preference SQL, which is translated to SQL, and several deployed applications.

\cite{BoKoSt01} introduces the {\em skyline} operator and describes
several evaluation methods for this operator.
As shown in Section \ref{sec:ext}, skyline  is a special case of winnow.
It is restricted to use a pure comparison ipf which is a conjunction
of pairwise comparisons of corresponding tuple components.
So in particular Example \ref{ex:wine}
does not fit in that framework. Some examples of possible rewritings for
skyline queries are given but no general rewriting rules are formulated.

\cite{AgWi00} uses quantitative 
preferences in queries and focuses on the issues arising in  combining such preferences.
\cite{HrKoPa01} explores in this context the problems of efficient query
processing. Since the preferences in this approach are based
on comparing the scores of individual tuples under given scoring
functions, they have to be intrinsic. 
However, the simulation
of extrinsic preferences using intrinsic ones (Section \ref{sec:ext})
is not readily available in this approach because the scoring functions
are not integrated with the query language. So, for instance, Example
\ref{ex:sal:6} cannot be handled.
In fact, even for preference relations that satisfy the property
of transitivity of the corresponding indifference relation, it is
not clear whether the scoring function capturing the preference
relation can be defined intrinsically (i.e., the function value
be determined solely by the the values of the tuple components).
The general construction of a scoring
function on the basis of a preference relation \cite{Fish99,Fish70}
does not provide such a definition.
So the exact expressive power of the quantitative approach to
preference queries remains unclear.

\subsection{Preferences in logic and artificial intelligence}
The papers on {\em preference logics} \cite{VWr63,Man91,Han01} address
the issue of capturing the common-sense meaning of preference
through appropriate axiomatizations. Preferences are defined on formulas,
not tuples, and with the exception of \cite{Man91,Cri02}
limited to the propositional case. 
\cite{Man91} proposes a modal logic of preference, and
\cite{Cri02} studies preferences in the context of relation algebras.
The application of the
results obtained in this area to database queries is unclear.

The papers on {\em preference reasoning} \cite{WeDo91,TaPe94,BoBrHoPo99} 
attempt to develop practical mechanisms for making inferences about
preferences and solving decision or configuration problems similar
to the one described in Example \ref{ex:wine}. 
A central notion there is that of {\em ceteris paribus}
preference: preferring one outcome to another, all else being equal. 
Typically, the problems addressed in this work are propositional
(or finite-domain).  Such problems can be encoded
in the relational data model and the inferences obtained by
evaluating preference queries. A detailed study of such an approach
remains still to be done. We note that the use of a full-fledged
query language in this context makes it possible to formulate 
considerably more complex decision and configuration problems
than before.

The work on {\em prioritized logic programming and nonmonotonic
reasoning} \cite{BrEi99,DeScTo00,SaIn00} has potential applications
to databases. However, like \cite{GoJaMa01} it relies on specialized
evaluation mechanisms, and the preferences considered are typically
limited to rule priorities.

\section{Conclusions and future work}\label{sec:concl}

We have presented a framework for specifying preferences
using logical formulas and its embedding into relational algebra.
As the result, preference queries and complex decision problems involving preferences
can be formulated in a simple and clean way.

Clearly, our framework is limited to applications that can be entirely
modeled within the relational model of data. 
Here are several examples that do not quite fit in this paradigm:
\begin{itemize}
\item preferences defined between {\em sets} of elements;
\item {\em heterogenous} preferences between tuples of different
arity or type (how to say I prefer a meal without a wine to a meal
with one in Example \ref{ex:wine}?);
\item preferences requiring nondeterministic choice. We believe
this is properly handled using a nondeterministic choice
\cite{GiGrSaZa97} or witness \cite{AbHuVi95} operator.
\end{itemize}

In addition to addressing the above limitations, future work
directions include:
\begin{itemize}
\item evaluation and optimization of preference queries, including cost-based
optimization;
\ignore{
incorporation of an operator that returns all the
tuples in a given relation sorted by rank (the ranking is defined
using the iterated winnow operator from  Section \ref{sec:iter});
}
\item extrinsic preferences;
\item defeasible and default preferences;
\item preference elicitation.
\end{itemize}

\section*{Acknowledgments}
This paper is dedicated to the memory of Javier Pinto whose premature
death prevented him from participating in this research.  The
conversations with Svet Braynov, Jarek Gryz, Bharat Jayaraman, and
Jorge Lobo, and the comments by the anonymous referees are gratefully
acknowledged. Special thanks go to Agnieszka Grabska for her skeptical
enthusiasm and timely feedback.

\end{document}